\documentstyle[]{laa}
\def\FIR{\ifmmode {\,$\tau_{\rm FIR}$} \else $\,\tau_{\rm FIR}$\fi}
\def\tk{\ifmmode {\,$T_{\rm k}$} \else $\,T_{\rm k}$\fi}
\def\mic{\ifmmode {\,\mu{\rm m}} \else $\,\mu{\rm m}$\fi}       
\def\kms{\ifmmode {\,{\rm km\,s^{-1}}}                          
	\else {\hbox{$\,$ {\rm km$\,$s$^{\rm -1}$}}}\fi}
\def\mo {\ifmmode {\,{\it M}\odot} \else $\,M$\odot\fi} 
\def\lo {\ifmmode {\,{\it L}\odot} \else $\,L$\odot\fi} 
\def\my {\ifmmode {\,{\it M}\solar\,{\rm yr^{-1}}}              
	\else {$\,M$\solar$\,$yr$^{\rm -1}$}\fi}
\def\cmm#1{\ifmmode {\,{\rm cm^{-#1}}}          
	\else \hbox{$\,${\rm cm$^{\rm -#1}$}}\fi}
\chardef\isp="10\def\i{\'\isp}		
\def\as {\ifmmode {^{\scriptscriptstyle\prime\prime}}           
	\else $^{\scriptscriptstyle\prime\prime}$\fi}
\def\am {\ifmmode {^{\scriptscriptstyle\prime}}                 
	\else $^{\scriptscriptstyle\prime}$\fi}
\def\deg {\ifmmode^\circ\else$^\circ$\fi}                       
\def\raw {\ifmmode\rightarrow\else$\rightarrow$\fi}             
\def\x {\ifmmode\times\else$\times$\fi}                         
\def\gsim {\ifmmode {\buildrel>\over\sim}               
	\else {\lower.6ex\hbox{$\buildrel>\over\sim$}}\fi}
\def\lsim {\ifmmode {\buildrel<\over\sim}               
	\else {\lower.6ex\hbox{$\buildrel<\over\sim$}}\fi}
\def\ra[#1 #2 #3.#4]{ #1$^{\rm h}$#2$^{\rm m}$#3$^{\rm s}$.#4}  
\def\dec[#1 #2 #3.#4]{ #1\deg#2\am#3{\as}.#4}             
\def\rax[#1 #2 #3]{RA: #1$^{\rm h}$#2$^{\rm m}$#3$^{\rm s}$}
\def\decx[#1 #2 #3]{Dec:#1\deg#2\am#3\as}          
\def\h2{\rm H$_2$}                      
\def\aua{{\rm A$\,$\&$\,$A}}            
\def\apj{{\rm ApJ}}                     
\def\apjs{{\rm ApJS}}                   

\begin{document}
\title{ISO SWS-LWS  observations of the prototypical reflection
nebula NGC 7023\footnotemark
}
\footnotetext{Based on observations with ISO, an ESA project
with instruments funded by ESA Member States (especially the PI
countries France, Germany, the Netherlands and the United Kingdom)
and with participation of ISAS and NASA}

\author{A. Fuente$^1$, J. Mart{\i}n-Pintado$^1$, 
N.J. Rodr{\i}guez-Fern\'andez$^1$, J. Cernicharo$^2$, 
M. Gerin$^3$}
\institute{$^1$ Observatorio Astron\'omico Nacional (IGN), Campus
 Universitario, Apdo. 1143, E-28800 Alcal\'a de Henares (Madrid),Spain\\
           $^2$ Instituto de Estructura de la Materia, 
Departamento de F{\i}sica Molecular, CSIC, Serrano 121, E-28006 Madrid, Spain\\
           $^3$ DEMIRM, Observatoire de Paris, 61 Av. de l'Observatoire, 
F-75014 Paris, France\\}

%
%
\thesaurus{09.01.1; 09.09.1 NGC 7023; 09.18.1; 08.09.2 HD 200775;
08.16.5 ; 13.09.4}
\maketitle
\begin{abstract}

We present SWS and LWS ISO observations towards a strip across
the photodissociation region (PDR) in the reflection nebula NGC 7023. 
SWS02 and LWS01 spectra have been 
taken towards the star and the brightest infrared filaments located
NW and SW from the star (hereafter NW and SW PDRs). In addition,
SWS02 spectra have been taken towards 
two intermediate positions (NW1 and SW1). This has provided 
important information about the extent and spatial distribution of the
warm H$_2$ and of the atomic species in this prototypical reflection nebula. 

Strong emission of the [SiII] 34.8$\mic$ line is detected towards the
star. While all the PDR tracers (the [CII] 157.7$\mic$, [OI] 63.2 and 
145.6$\mic$,
[HI] 21cm and the H$_2$ rotational lines) present a ring-like morphology
with the peaks toward the NW and SW PDRs and a minimum around the star,
the SiII emission is filling the hole of this ring with
the peak towards the star. 
This morphology can only be explained if the SiII emission arise in the 
lowest extinction layers of the PDR (A$_v$ $<$ 2 mag) and the HII region.
At least 20\% -- 30\% of the  Si must be in gas phase in these layers.
For A$_v$ $\geq$ 2 mag, the Si is mainly in solid form ($\delta$ Si = -1.3).  

The NW and SW PDRs have very similar excitation conditions, high density
filaments (n$\sim$ 10$^6$ cm$^{-3}$) immersed in a more diffuse
interfilament medium (n$\sim$ 10$^4$ cm$^{-3}$). In both, 
the NW and SW PDRs, the 
intensities of the H$_2$ rotational lines can only be fitted by
assuming an ortho-to-para-H$_2$ ratio lower than 3 in gas with
rotation temperatures from 400 to 700 K. Therefore, there is a
non-equilibrium OTP ratio in the region. Furthermore, the comparison
between the OTP ratio derived from H$_2$ vibrational lines and
the pure H$_2$ rotational lines, shows that the OTP ratio increases 
from $\sim$ 1.5 to 3 across the photodissociation region with
larger values in the less shielded gas (A$_v$$<$ 0.7 mag).
This behavior is interpreted as a consequence of an advancing 
photodissociation front.
  
We have not detected the OH, CH and CH$_2$ lines towards the observed
positions. This is consistent with the weakness of these lines in
other sources and can be explained as a consequence of the small beam
filling factor of the dense gas in the LWS aperture. 
The CO J=17$\rightarrow$16 line
has been tentatively detected towards the star.   
\end{abstract}


\section{Introduction}
NGC 7023 is a prototypical photodissociation region that
has been largely studied in the last 10 years. It is a reflection
nebula illuminated by the Herbig B3Ve star HD 200775. HD 200775 is located 
in a cavity of the molecular cloud whose sharp edges delineate
perfectly the optical nebula (Fuente et al. 1992, Rogers et al. 1995,
Fuente et al. 1998). 
Observations of CI and OII by Chokshi et al. (1988)
reveal that a dense PDR is formed in the walls of this cavity
with its peak located 50'' NW from the star. 
Chokshi et al. (1988) carried out the first model for this region and 
estimated a UV field of 
G$_0$($\lambda$$>$ 912 \AA)=2.4 10$^3$ in units 
of the Habing field and a density
of n = 10$^4$ cm$^{-3}$. Because of its edge-on geometry 
and proximity (d $\sim$ 440 pc), this PDR
turned out to be one of the best
sites to study the physical and chemical processes taking place in
a PDR. Fig. 1 shows the column density map of HI as derived from
the [HI] 21 cm line superposed on the integrated intensity map of the
$^{13}$CO J=1$\rightarrow$0 line. The HI peak is 
shifted $\approx$ 20 $''$
relative to the molecular ridge, showing 
the layered structure expected in a PDR (Fuente et al. 1998).

Very high angular resolution images of the region in the Extended Red
Emission (ERE) and vibrationally excited H$_2$ lines show that the PDR 
has a filamentary
structure with very bright thin filaments located 50$''$ NW and 70$''$ SW
from the star (Sellgren et al. 1992, Lemaire et al. 1996). 
Hereafter, we will refer to these positions as NW PDR
and SW PDR respectively. The NW PDR is spatially coincident with the
peak of the CI and OII lines and has been
extensively studied in atomic and molecular lines (Chokshi et al. 1988;
Fuente et al. 1993, 1996, 1997; Lemaire et al. 1996,1999; 
Martini et al. 1997; Gerin et al. 1998).
Fuente et al. (1993, 1997) observed that
the CN/HCN and CO$^+$/HCO$^+$ abundance ratios 
increases by a factor 10 and $>$ 100 respectively towards this position
relative to the values in the molecular cloud.
Both, the CN/HCN
and CO$^+$/HCO$^+$ abundance ratios are expected to increase in the
lower extinction layers of PDRs (Fuente et al. 1993,
Sternberg \& Dalgarno 1995). 
An interferometric image of the HCO$^+$ line towards the NW PDR
show that the PDR has also
a filamentary structure in molecular emission
with high density filaments (n $>$ 10$^5$ cm$^{-3}$) embedded in a 
more diffuse medium (n $\sim$ 10$^4$ cm$^{-3}$) (Fuente et al., 1996). 
The HCO$^+$ filaments are spatially 
coincident with the filaments seen in the near-infrared continuum 
images and in the emission of the vibrational H$_2$ lines.

The SW PDR is fainter by a factor 2-3 than the  northern one and has been less
studied in both, molecular and atomic lines. Gerin et al. (1998) reported 
observations of this PDR in CO and CI lines.   

We have observed a strip 
which joins the NW PDR, the star and SW PDR using the SWS and LWS 
instrument on board of ISO. These observations have provided a very
important information about the extent and spatial distribution of the
warm H$_2$ and the atomic species within the PDR.

\section{Observations}

The observations were made using the Long- Wavelength Spectrometer 
(LWS) (Clegg et al. 1996, Swinyard et al. 1996) and the 
Short-Wavelength Spectrometer (SWS) (de Graauw et al. 1996) on 
board the Infrared Space Observatory (ISO) (Kessler et al. 1996).
The LWS observations were carried out in full grating scan mode (LWS01 AOT)
which provides coverage of the 43 - 90.5 $\mic$ range with a 
spectral resolution of 0.29 $\mic$ and of the 90.5 - 196.5 $\mic$
range with a resolution of 0.60 $\mic$ with ten independent detectors.
Since the instrumental beam size is 74$''$ - 90$''$, we have observed
only three positions of the nebula, the NW PDR, the star and the SW PDR.
In Table 1, we give the coordinates of 
the observed positions. The data were processed using version 6 of
the off-line pipeline (OLP V6). The  uncertainty in the calibration is 
of about 30 \% (Swinyard et al. 1996)
 
We have observed one HI recombination line (Br$\alpha$),
five H$_2$ pure rotational lines (S(0),S(1),S(2),S(3),
S(4) and S(5) v=0--0) and
four ionic and atomic fine structure lines 
([SIV] 10.5 $\mic$, [FeII] 26.0 $\mic$, [SIII] 33.5 $\mic$ and 
[SiII] 34.8 $\mic$) using SWS AOT02. This mode
provides a spectral resolution of $\lambda$/$\Delta$$\lambda$ $\sim$ 1000-2000.
All the observed lines are unresolved at this spectral resolution. The
aperture of the instrument in these line ranges from 14$''$ x 20$''$ to
20$''$ x 33$''$. In order to study the spatial distribution of these lines
we have observed 5 positions along the strip that joins the
NW PDR, the star and the SW PDR. 
These positions are marked in Fig. 1, the observed spectra are shown in
Fig. 2 and absolute coordinates  
are given in Table 1.
Data reduction are carried out with version 7.0 of the Off Line
Processing routines and the SWS Interactive Analysis at the ISO Spectrometer
Data Center at the Max-Planck Institut f\"ur Extraterrestrische Physik. 
The  uncertainties in the calibration are of about 20 \% at
short wavelengths (2 - 20 $\mic$) and of about 30\% at longer wavelengths 
(Salama et al. 1997).

\begin{table*}[t]
\caption{Observations}
\begin{tabular}{l c }
\\ \hline
\multicolumn{1}{l}{Position} &
\multicolumn{1}{c}{Coordinates (2000)}  
  \\  \hline
NW PDR & \ra[21 01 32.5] \dec[68 10 27.5]   \\
NW1    & \ra[21 01 34.5] \dec[68 10 05.0]   \\
Star   & \ra[21 01 36.9] \dec[68 09 48.3]  \\
SW1    & \ra[21 01 36.9] \dec[68 09 10.0]    \\
SW PDR & \ra[21 01 32.3] \dec[68 08 45.0]   \\
\hline
\end{tabular}
\end{table*}

\section{Results}

The LWS and SWS observational results are shown in Table 2 and 3.
The spectra of the SWS observations are shown in Fig. 2.
In order to study the spatial distribution
of the different lines, we have plotted in Fig. 3 the normalized 
intensities of the [HI] 21cm line (Fuente et al. 1996,1998), 
the H$_2$ S(1) v=0--0 and [SiII] 34.8 $\mic$ lines
as a function of the distance from the star.

\subsection{Recombination lines}
The Br$\alpha$ line 
has only been detected towards the star. Continuum emission
at 3.6 cm and  6 cm was detected by Skinner et al. (1993).
They interpreted the emission as arising in an ionized anisotropic wind.
Later, Nisini et al (1995) observed the Pa$\beta$, 
Br$\gamma$, Br$\alpha$, Pf$\gamma$ and Pf$\beta$ recombination lines with
an aperture of 10$''$ and modeled the wind with a mass loss rate of
3.9 10$^{-7}$ M$_\odot$ yr$^{-1}$ and a terminal 
velocity of 280 kms$^{-1}$. 
We have measured an intensity of the Br$\alpha$ line of 
1.4 10$^{-3}$ erg s$^{-1}$ cm$^{-2}$ sr$^{-1}$  which agrees
within the calibration  uncertainties with the data of
Nisini et al. (1995). 

\subsection{The CII and OI lines}
The [CII] 157.7 $\mic$ and the [OI] 63 and 145.6 $\mic$ lines 
have been detected towards all the observed positions. The intensities 
and line intensity ratios are shown in Table 2.
The emission of the three lines is maximum in 
the NW PDR and decreases towards
the south, being the SW PDR fainter by a factor of 2 than the NW PDR. 
Along the observed strip, the OI(63$\mic$)/OI(145.6$\mic$) 
and  OI(63$\mic$)/CII(157.7$\mic$) ratios are uniform and
take the values $\sim$ 10 and $\sim$ 2 respectively. This means that
at the scale of the LWS observations, the physical conditions of the
nebula are quite uniform. High angular resolution observations show that
this PDR is formed by high density filaments immersed in a more diffuse
medium (Sellgren et al. 1992, Lemaire et al. 1996, Fuente et al. 1996). 
Because of the low angular resolution of the LWS observations,
they are very likely tracing the more extended diffuse medium.

\subsection{Other ionic and atomic fine structure lines: SiII}
The spatial distribution of the SiII line is very different
from that of the CII, OI and HI lines (see Fig. 3). 
While the CII, OI and HI lines 
peak towards the NW PDR, the SiII line
peaks towards the star, with the intensity towards the
star $\sim$ 3 times larger than towards the NW PDR. This 
different spatial distribution cannot be explained by the different
angular resolution of the CII and OI observations.
The [HI] 21cm line map which has an angular resolution similar to that of the
SiII line presents a ring-like morphology with the peaks towards
the NW and SW PDRs and a minimum towards the star, just in the
peak of the SiII line. 
The implications of the peculiar spatial distribution of the SiII emission
are discussed in Section 4.1. 
We have not detected the [SIII] 33.480 $\mic$, [SIV] 10.51 $\mic$
and [FeII] 25.9$\mic$ lines at any position (see Table 3 for upper limits). 

\subsection{H$_2$ rotational lines}
We have observed the v=0--0 H$_2$ rotational lines from S(0) to S(5) towards
the positions shown in Fig. 1. The intensities of the H$_2$ lines for
all positions are shown in Table 2.
Towards the positions NW1, SW1 and the star, 
we have detected only the S(1) line 
which prevents from any excitation analysis. 
The most intense H$_2$ emission
is found towards the NW and SW PDRs. These positions are spatially coincident 
with the filaments observed in H$_2$ fluorescent emission which are located
in the interface between the atomic and the molecular gas. The H$_2$ emission
is much weaker towards the HI clump (position NW1) and the SW1 position, 
suggesting that this region is mainly atomic. 
In contrast with the spatial distribution of HI which
is more intense towards NW1 than towards SW1, 
the spatial distribution of the H$_2$ emission is quite symmetric around
the star.

\subsection{Molecular lines}

Several molecular lines of species that are expected to be abundant in PDRs
(CO,OH,CH,CH$_2$,H$_2$O) have observable lines in the wavelength range covered
by the LWS01 spectrum. In the LWS01 spectrum towards NW PDR we have 
tentatively 
detected a weak line centered at 153.927 $\mic$ with a S/N ratio 
of $\sim$ 3.
A weak line centered at 153.637 $\mic$ with a S/N ratio of 5
appears also in the spectrum towards the star. Both spectra are shown
in Fig. 4. Taking into account that
the spectral resolution of the LWS at these frequencies is 0.6$\mic$, 
both lines can be identified with the CO J = 17$\rightarrow$16 line 
(153.267 $\mic$). However, since other CO lines have not been detected towards
these positions, we cannot discard the possibility of a misidentification or
a spurious detection. 
Towards the SW PDR, we have not detected any molecular line with  
typical upper limits of 
$\sim$ 5 10$^{-12}$ - 5 10$^{-13}$ erg s$^{-1}$ cm$^{-2}$ $\mic$$^{-1}$ 
depending on the detector being the best limits for detectors 8 to 9 
(161.2 to 196.7 $\mic$).

\begin{table*}[t]
\caption{Intensitites of the OI and CII lines}
\begin{tabular}{l ccc }
\\ \hline
\multicolumn{1}{l}{} &
\multicolumn{1}{c}{NW PDR} & 
\multicolumn{1}{c}{star} &
\multicolumn{1}{c}{SW PDR}  \\ 
( x 10$^{-4}$ erg s$^{-1}$ cm$^{-2}$ sr$^{-1}$)  \\ \hline
OI(63$\mic$)  & 7.5$\pm$0.1 & 3.9$\pm$0.8  & 3.8$\pm$1.3  \\
OI(145.6 $\mic$) & 0.9$\pm$0.1 & 0.4$\pm$0.1 & 0.3$\pm$0.1  \\
CII(157.7 $\mic$) & 2.8$\pm$0.1 & 1.8$\pm$0.1 & 1.7$\pm$0.1 \\
OI(63$\mic$)/OI(145.6 $\mic$) & 8.3 & 9.0 & 12.7  \\
OI(63$\mic$)/CII(157.7 $\mic$) & 2.7 & 2.2 & 2.3  \\ 
\hline
%
\end{tabular}

\caption{SWS observations}
\begin{tabular}{l cccccc }
\\ \hline
\multicolumn{1}{l}{Line} &
\multicolumn{1}{l}{$\lambda$($\mic$)} &
\multicolumn{1}{c}{NW PDR} &
\multicolumn{1}{c}{NW1} &
\multicolumn{1}{c}{Star} &
\multicolumn{1}{c}{SW1} &
\multicolumn{1}{c}{SW PDR}  \\ 
(x 10$^{-5}$ erg s$^{-1}$ cm$^{-2}$ sr$^{-1}$)  \\ \hline
\multicolumn{7}{c}{Recombination lines} \\\hline
Br$\alpha$ & 4.051 & $\leq$2.3 & $\leq$1.0 & 141.7$\pm$1.5 & $\leq$0.7 &
$\leq$1.0  \\ \hline
\multicolumn{7}{c}{H$_2$ rotational lines} \\\hline
H$_2$ S(5) & 6.909 & 26.4$\pm$3.0 & $\leq$4.8 & $\leq$7.1 & $\leq$3.3 &
9.2$\pm$3.3 \\
H$_2$ S(4) & 8.025 & 15.0$\pm$0.2 & $\leq$3.9 & $\leq$5.0 & $\leq$1.4 &
5.6$\pm$1.1  \\
H$_2$ S(3) & 9.665 & 40.8$\pm$1.3 & $\leq$4.4 & $\leq$3.3 & $\leq$1.6 &
21.2$\pm$1.5  \\
H$_2$ S(2) & 12.279 & 24.0$\pm$3.3 & $\leq$10 & $\leq$7.4 & $\leq$8.0 &
16.5$\pm$2.7  \\  
H$_2$ S(1) & 17.035 & 21.4$\pm$0.7 & 1.7$\pm$0.4 & 1.8$\pm$0.2 & 
1.6$\pm$0.4 & 10.0$\pm$0.7  \\
H$_2$ S(0) & 28.219 & 4.8$\pm$4.8 & $\leq$4.4 & $\leq$3.5 & $\leq$3.3 &
$\leq$4.5  \\ \hline
\multicolumn{7}{c}{Atomic Fine Structure Lines} \\\hline
SIV & 10.510 & $\leq$2.3 & $\leq$2.9 & $\leq$3.1 & $\leq$ 1.8 & $\leq$1.7 \\
FeII & 25.988 & $\leq$2.3 & $\leq$2.0 & $\leq$1.4 & $\leq$2.1 & $\leq$2.2 \\
SIII & 33.480 & $\leq$3.3 & $\leq$6.6 & $\leq$8.3 & $\leq$1.7 & $\leq$3.3 \\
SiII & 34.814 & 3.4$\pm$1.3 & 4.7$\pm$0.5 & 8.6$\pm$0.1 & 1.8$\pm$0.3 & 
$\leq$3.4 \\
\hline
\end{tabular}

\end{table*}

\section{Discussion}

The NW PDR was modeled for the first time by Chokshi et al. (1988)
with an incident UV field of G$_0$=2.4 10$^3$ and a density
of n = 10$^4$ cm$^{-3}$. Models of this region based on
H$_2$ ro-vibrational lines (Martini et al. 1997), CO, CI and CII lines
(Gerin et al. 1998) and H$_2$ rotational lines (Fuente et al. 1999)
give slightly higher values of the UV field, G$_0$= 10$^4$, and
density, n = 10$^5$ --  10$^6$ cm$^{-3}$. We can estimate
the UV field and density of the PDR based on the observed
OI and CII lines.
Along the observed
strip, the OI(63$\mic$)/OI(145.6$\mic$) ratio is $\sim$ 10.
This value is lower by a factor $\sim$ 2-3 than those predicted by PDR models
for a region with a UV field of 10$^3$ and a density of 
$\sim$ 10$^4$ cm$^{-3}$ (see e.g. Kaufman et al. 1999, Burton et al. 1992 and
references therein).  
Low OI(63$\mic$)/OI(145.6$\mic$) ratios
have been observed towards other PDRs and interpreted as the
consequence of the self-absorption of the 63 $\mic$ line 
by the foreground gas (Lorenzetti et al. 1999). 
This interpretation
is not plausible for our region in which the extinction by the foreground
dust is $<$ 2 mag (Fuente et al. 1998, Gerin et al. 1998). 
We have no explanation for the low observed values of 
the OI(63$\mic$)/OI(145.6$\mic$) ratio.
 
Because the [OI] 63$\mic$ and [CII] 157.7$\mic$ lines
have different critical densities, 
the OI(63$\mic$)/CII(157.7$\mic$) ratio constitute a 
good density tracer for densities 10$^3$ - 10$^6$ cm$^{-3}$.
The value of the OI(63$\mic$)/CII(157.7$\mic$) is $\sim$ 2 
towards all the observed positions. This value is 
consistent with model predictions for G$_0$ $\sim$ 10$^3$ and 
n $\sim$ 10$^3$ -- 10$^4$ cm$^{-3}$ which are the physical parameters
derived by Chokshi et al. (1988) from the same lines and 
better angular resolution observations. 
Densities around 10$^4$ cm$^{-3}$ are
also derived from the rotational lines of CO and its isotopes 
(Fuente et al. 1993, Rogers et al. 1995, Gerin et al. 1998) and from 
the VLA HI observations (Fuente et al. 1993,1998). 
However, densities as high as 
n = 10$^6$ cm$^{-3}$ are derived from the H$_2$ ro-vibrational
and rotational lines (Lemaire et al. 1996, Martini et al. 1997,
Fuente et al. 1999, this work). 
High densities (n = 10$^5$ cm$^{-3}$) are also derived from
the observations of molecular lines towards the NW PDR (Fuente et al.
1993, 1996).  
As commented in Section 3.2, this PDR is formed by high density filaments
immersed in a more diffuse medium. The observed OI(63$\mic$)/CII(157.7$\mic$) 
ratio is very likely determined by the value of this ratio in the
diffuse component, n $\sim$ 10$^4$ cm$^{-3}$, which fills most
of the LWS aperture. In fact, the better angular resolution 
[CII] 157.7 $\mic$ and [OI] 63 $\mic$
data reported by Chokshi et al. (1988) and Gerin et al. (1998) 
show that the emission of both lines is extended over the whole HI region.  
The high density filaments (n $>$ 10$^5$ cm$^{-3}$)
are better traced by the observations of
the H$_2$ rotational and ro-vibrational lines towards the NW and SW PDRs, 
as well as 
by the observation of the rotational lines of molecules with
high dipole moment.

\subsection{SiII}

The emission of the [SiII] 34.8 $\mic$ line 
is considered to arise mainly in the
neutral part of the cloud, i.e., in the PDR. 
Silicon has a first ionization potential of 8.15 eV and
remains singly ionized until an optical depth of A$_v$$\sim$ 5, 
a value which is fairly independent of n and G$_0$.
With a critical density of n = 3.4 10$^5$ cm$^{-3}$, 
the intensity of the [SiII] 34.8 $\mic$ 
line is strongly dependent on the uncertain Si abundance in gas phase. 
Silicon is heavily depleted in the ISM with logarithmic
depletion values ($\delta_{Si}$ = log(X$_{Si}^{gas}$/X$_{Si}$)) 
ranging from -1.30 to -0.18. 
Most PDR models adopt an Si abundance of 3.6 10$^{-6}$ ($\delta_{Si}$= -1.0)
which is the mean value derived by van Steenberg \& Shull (1988) but the
actual value in the PDR could differ by an order of magnitude, 
and consequently the intensity of the [SiII] 34.8 $\mic$ line. 
The [SiII] 34.8 $\mic$ line is usually compared with 
the [OI] 63 $\mic$ line because both lines have similar excitation conditions
and their total column densities are fairly independent 
of n and G$_0$. Standard PDR models ($\delta_{Si}$= -1.0) predict  
SiII(34.8$\mic$)/OI(63$\mic$) intensity 
ratios ranging from  $\sim$ 0.06 - 0.08 for G$_0$ = 10$^3$ - 10$^4$ and
densities $\sim$ 10$^4$ - 10$^5$ cm$^{-3}$ (e.g., Burton et al. 1992). 

As it is shown in Fig. 2 and 3, the spatial
distribution of the SiII emission
in NGC 7023 is different from that of all the other PDR tracers.
All the PDR tracers peak towards the vibrationally excited 
H$_2$ filaments located in the walls of the cavity in which the star
is immersed (the NW and SW PDRs) while the SiII emission
fills this cavity and peaks towards the star. 
The intense H$_2$ filaments observed towards 
the NW and SW PDRs show that the 
extinction from the star to these positions is lower than 2 mag. 
Then, the SiII emission is arising in the gas located at 
an extinction of $<$ 2 mag from the star. 
 
We have compared the SiII data with the OI observations 
reported by Chokshi et al. (1988) which have a FWHM aperture of 33$''$. 
Chokshi et al. (1988)
measured an intensity of 1.2 $\pm$ 0.1 10$^{-3}$ erg s$^{-1}$ cm$^2$ sr$^{-1}$
towards the NW PDR. The SiII(34.8$\mic$)/OI(63$\mic$) intensity 
ratio is 0.03 at this position. This value is in agreement
with PDR models if the assumed Si elemental abundance is
$\sim$ 1.5 10$^{-6}$ ($\delta_{Si}$= -1.3), i.e., silicon is heavily
depleted with only the 5\% in gas phase .      
Chokshi et al. (1988) observed a position shifted  20$''$ W from the
star and measured an intensity of 6.2$\pm$0.1 10$^{-4}$ erg cm$^{-2}$ 
s$^{-1}$ sr$^{-1}$. 
Taking into account the dust and gas
distribution around the star (see Fig. 1), this value can be
considered an upper limit to the
actual OI emission and we can conclude that 
the SiII (34.8$\mic$)/OI (63$\mic$) line 
ratio is $>$ 0.14 towards the star, i.e. a factor of 4 larger 
than towards the NW PDR.  
If the SiII emission arise mainly in neutral gas, the Si abundance 
in gas phase
must be larger by more than a factor of 4 towards the star than towards
the NW PDR, i.e. at least 20\% of the silicon is in gas phase towards the
star.    

Even more suggestive is to compare the SiII with HI emission. The 
SiII/HI ratio presents a symmetric distribution around the star, in
which the SiII/HI ratio increases towards the star and is  
a factor of 7 larger towards
the star than towards the NW and SW PDRs. If the SiII arise in the
mainly neutral layer traced by HI, more than 30\% of the Si is in
gas phase towards the star.     
Other possibility is that an important fraction of the SiII emission
arise in ionized gas which is not traced by the HI emission. 
The contribution of the ionized wind to the
intensity of the SiII line is expected to be negligible. 
We have carried out some model calculations using the
code CLOUDY\footnotemark (Ferland, 1996) 
assuming the Kurucz stellar atmosphere for a star with
T$_{eff}$ = 17000 K, log g=4 cm$^{-2}$ (Kurucz 1979) and the wind parameters
derived by Nisini et al. (1995).
These calculations show that even assuming
the solar abundance for silicon, the SiII intensity in the
ionized wind is expected to be a factor of 10$^{-7}$ lower than
the intensity of the Br$\alpha$ line.  Alternatively,
the SiII emission could arise in a small HII region created by the
star. The  SiII intensity in an ionized sphere around the star 
with constant density, n = 10$^2$ cm$^{-3}$, 
could be as high as 3.2 10$^{-5}$ erg s$^{-1}$ cm$^{-2}$ sr$^{-1}$
if all the silicon is in gas phase, i.e.,
$\approx$ 40\% of the SiII emission towards the star. In this case
only the 60\% of the emission observed towards the star would arise
in the neutral layer traced by HI. But 
even in this case the Si abundance in the neutral gas must increase
by a factor of 4 to explain the SiII/HI ratio.

\footnotetext{We have used the version number 90.04 which uses the
MICE interface developed by Henrik Spoon at MPE }

Then we can conclude that silicon is heavily depleted for 
A$_v$$\geq$ 2 mag ($\delta_{Si}$= -1.3) and its abundance 
increases progressively towards the star where at least 20\% -- 30\% 
of the silicon must be in gas phase.
   
At this point, it is interesting to consider the case of the 
[FeII] 26.0$\mic$ line. We have not detected the emission of
the [FeII] 26.0$\mic$ line towards any position.
This lack of emission
can be explained by excitation effects and a different depletion
of Fe and Si on grains. The critical density of the 
[FeII] 26.0$\mic$ line is an order of magnitude larger than that
of the [SiII] 34.8$\mic$ line (Hollenbach \& Mc Kee 1979). 
On the other hand, Si is more easily released
to the gas phase than Fe. Fe is expected to be
significantly depleted in the HII region and the PDR 
although the Si has been 
released to the gas phase (Sofia et al. 1994).
In standard PDR models, the depletion of Fe is assumed to be a factor of 
10 larger than that of Si ($\delta_{Fe}$= -2.0) and    
the intensity of the FeII(26.0$\mic$)
lines is then an order of magnitude lower than the intensity of the
SiII (34.8$\mic$) line 
(Burton et al. 1992). 

It is well known that the abundance of gas phase silicon in the
interstellar medium increases by grain-grain collisions and/or 
sputtering in shocks (Mart{\i}n-Pintado et al. 1992,
Caselli et al. 1997, Bachiller \& P\'erez-Gutierrez 1997). 
Photodesorption of Si in grain mantles by UV radiation 
has recently been proposed as a mechanism to explain 
the large SiO abundance in some PDRs (Walmsley et al. 1999). 
Our data put some constraints on the efficiency of photodesorption
to release Si to the gas phase. In NGC 7023, Si is heavily depleted
($\delta_{Si}$= -1.3) in gas at an extinction of 2 mag, i.e., silicon
is mainly in solid form in the molecular gas where the emission of
SiO is expected to arise. 
The large bipolar cavity associated with HD 200775 proves the existence
of an energetic bipolar outflow in a previous stage of the
stellar evolution (Fuente et al. 1998). The large gas phase Si abundance
in the positions closest to the star could be a relique of this stage.
If it were the case, the efficiency of photodesorption to return the
Si to the gas phase would be even lower.

\subsection{A gradient in the ortho-to-para-H$_2$ ratio}
The S(0),S(1),S(2), S(3), S(4) and S(5) v=0--0 H$_2$ rotational lines have 
been detected towards the NW PDR  and the S(1),S(2), S(3), S(4) and S(5) 
lines towards the SW PDR. In a previous paper (Fuente et al. 1999), 
we have studied in detail the H$_2$ rotational lines 
towards the NW PDR. In this paper we will concentrate in the observations 
towards the SW PDR. 

In Fig. 5 we show the rotational diagram of the H$_2$ lines towards the
SW PDR. Note that the errors in Fig. 5 are entirely dominated by the
calibration  uncertainties (15\% at 6.909 $\mic$, 25\% at 8.025 $\mic$,
25\% at 9.665 $\mic$, 25\% at 12.279 \%, 20\% at 17.035 $\mic$ and 
30\% at 28.219 $\mic$).
The data have been corrected for dust attenuation using the
value for the dust opacity derived from the LWS01 spectra 
(A$_v$ = 0.3 mag) and the
extinction curve of Draine \& Lee (1984). The extinction for the
S(0), S(1), S(2), S(3), S(4) and S(5) lines amounts 
to 0.006,0.013, 0.013, 0.027,0.013 and 0.006.
Like in the case of the NW PDR (see Fuente et al. 1999), 
the rotational diagram of the SW PDR shows that the ortho-H$_2$ levels
have systematically lower $N_u/g_u$ values (where $N_u$ and $g_u$ are the 
column densities and degeneracies of the upper levels of the transitions)
than the adjacent J-1 and J+1 para-H$_2$ levels producing
a ``zig-zag" distribution. In fact, the ortho-levels seem to define a
curve which is offset from that of the para-levels by more than a
factor of 2 (see Fig. 5). This
offset cannot be explained by calibration uncertainties and extinction
effects. The calibration uncertainties are at most 30\%. 
Since the values of the extinction at 17.03 $\mic$, 12.3 $\mic$ 
and 8.025 $\mic$
are very similar (Draine \& Lee 1984), a correction for dust attenuation
cannot push the S(1), S(2) and S(4) points to the same curve.  
This "zig-zag'' distribution cannot be due to the different apertures
of the instrument for the different lines. In the case of a point-like
source, the S(1) and S(2) lines should be corrected by a factor 1.35 relative
to the S(3) line and the offset between ortho- and para- curve would increase.
 
The "zig-zag'' distribution observed in the rotational diagram
can only be fitted by assuming a non-equilibrium OTP ratio, i.e.,
different from the OTP equilibrium value 
at the gas kinetic temperature.
In this case, only the rotation temperature 
between the levels of the same
symmetry constitute an estimate of the gas kinetic temperature.     
Using the ortho-levels, we have
derived a rotation temperature of $\sim$ 450 K between
the S(1) and S(3) lines and  of $\sim$ 630 K between the S(3) and S(5) lines.
For the para-levels we have derived a rotation temperature of $\sim$ 430 K
using the S(2) and S(4) lines. Then we can conclude that the H$_2$ rotational
lines are arising in gas with kinetic temperatures ranging from 
$\sim$ 400 to 700 K. These kinetic temperatures are 
similar to those obtained in the NW PDR, but the intensities of the 
lines are about a factor of 2 lower. A change in the incident
UV field and/or the density would imply a variation of
both the line ratios (rotation temperatures) and the line intensities. 
Since the line ratios are similar to those of the
NW PDR and the line intensities are significantly lower, we conclude
that the SW PDR has the same excitation conditions 
as the NW PDR (G$_0$ = 10$^4$, n = 10$^6$ cm$^{-3}$)
but a beam filling factor of $\sim$ 0.5. 
All PDR models which include both, chemical and thermal balance
in the gas,
assume a fixed value of 3 for the OTP ratio, which is the
equilibrium value for gas kinetic temperatures $>$ 100 K. 
In order to determine the value of the OTP ratio in the SW PDR,
we have corrected the intensities predicted by
the model of Burton et al. (1992) for the effect of different
values of the OTP ratio assuming that the total amount of
H$_2$ molecules at each temperature and the line ratios between
the levels of the same symmetry remain unchanged.
This correction is only valid if the excitation of the H$_2$ lines
is mainly collisional which is the expected case for the low
rotational transitions. In Fig. 6, we compare     
the observational data for the S(0), S(1), S(2) and S(3) transitions
with the model for G$_0$ = 10$^4$, n = 10$^6$ cm$^{-3}$ and a fixed
OTP ratio of 3, and
the same model corrected for OTP ratios of 1.5 and 2. 
We obtain that the rotation diagram of the SW PDR 
is well fitted with an OTP ratio of 1.5 
(see Fig. 6).
Martini et al. (1997) derived an 
OTP ratio of 2.3$\pm$0.5 towards the SW PDR
based on the H$_2$ vibrational lines. Because of the optical depth effects in 
the excitation of the ortho- and para- H$_2$ vibrational lines, 
this value is just a lower limit to the actual value of the OTP ratio
(Draine \& Bertoldi 1996, Sternberg \& Neufeld 1999). 
Then we can conclude that the OTP ratio is
larger than 2.3$\pm$0.5 in the less shielded layers of the
PDR (A$_v$ $<$ 0.7 mag) where the vibrational lines arises. 
The gas kinetic temperature in these layers ranges 
from several hundreds to 2000 K. 
Comparing the OTP ratio derived from the rotational lines with that derived
from the vibrational lines, we conclude that the OTP ratio 
increases across the PDR. It takes values $\sim$ 1.5 at temperatures 
of $\sim$ 400 - 700 K and values close to 3 at 
temperatures $\gsim$ 700 K. 

The same behavior, a non-equilibrium OTP ratio for the gas with
kinetic temperatures $\sim$ 400 K and a gradient of the OTP ratio 
across the PDR was found in the NW PDR by 
Fuente et al. (1999). They discussed that the most likely
explanation for this behavior is the existence of an advancing 
photodissociation front. An important problem remains in this
explanation, the velocity of the photodissociation front required to have 
a non-equilibrium OTP ratio is too large  ($\sim 10^7 n^{-1}$ kms$^{-1}$).
However, since the expected behavior of the OTP ratio in an 
advancing photodissociation front is in agreement with that observed
in our data (the cooler gas is further from the OTP equilibrium 
value than the warm gas)  
and there are many uncertainties in this estimate 
(it is not known the OTP ratio at which the H$_2$ is formed within the PDR, 
the gas is clumpy and small dense clumps can be evaporating into the PDR,...), 
we considered that this was the
most likely explanation. Later,   
Lemaire et al. (1999) have obtained a very high spectral 
resolution spectrum of the H$_2$ S(1) v=1$\rightarrow$0 line at 2.121 $\mic$
towards the NW PDR. The H$_2$ line is seen shifted in velocity 
relative to that of the HI and the CO rotational lines. 
They argue that this velocity difference is 
a clear proof of the presence of dynamical effects in this PDR and
estimated a (projected) velocity of 1 kms$^{-1}$ for the photodissociation
front. So far, very few models have been developed for non-equilibrium
PDRs (Bertoldi \& Draine 1996, St\"orzer \& Hollenbach 1998).
These models consider regions illuminated by O
stars in which a ionization front is combined with a photodissociation front
and assume a fixed value for the OTP ratio of 3. They do not give 
any information about the effect of an advancing photodissociation on the
ortho-to-para- H$_2$ ratio. But they predict how the existence of a  
photodissociation front affects the thermal structure of
the region. Following the model by St\"orzer \& Hollenbach (1998),
because of the advection of molecular gas through the PDR, 
the intensities of the H$_2$
rotational lines and vibrational lines can be enhanced by a factor of 3.
Assuming  a velocity of 1 kms$^{-1}$ for the photodissociation front, 
the intensities of the H$_2$ rotational lines observed in the NW PDR
can be well fitted with a density of n$\sim$ 10$^5$ cm$^{-1}$ and a UV field
of G$_0$ = 10$^4$. A non-equilibrium OTP ratio has to be assumed
to explain the "zig-zag'' distribution.

\subsection{Absence of molecular emission}
Towards NW PDR we have tentatively 
detected a weak line centered at 153.927 $\mic$ with a S/N ratio of $\sim$ 3.
A weak line centered at 153.637 $\mic$ appears also in the LWS spectrum
towards the star position with a S/N ratio of 5. Taking into account that
the spectral resolution of the LWS at these frequencies is 0.6$\mic$, 
both lines can be identified with the CO J = 17$\rightarrow$16 line 
(153.267 $\mic$). However, since other CO lines have not been detected towards
these positions, we cannot discard the possibility of a misidentification or
a spurious detection. If confirmed, this line would implied the existence of
a gas with T$_k$$>$ 200 K and n$>$ 10$^5$ cm$^{-3}$ in the vicinity of 
the star. However the density of the bulk of the molecular gas in the PDR
must be at least an order of magnitude lower. 
Federman et al. (1997) estimated based on the ultraviolet absorption lines
of CH, CH$^+$, C$_2$ and CN that the density of the PDR in front of the star
is $\sim$ 200 cm$^{-3}$. Furthermore,
the OI(63$\mic$)/CII(157.7$\mic$) ratio is consistent with densities of
10$^4$ cm$^{-3}$. 

We have not detected any OH, CH and CH$_2$ line towards the observed
positions.
Emission of the OH lines have been detected
in several sources (NGC 7027: Liu et al. 1996; R CrA and LkH$\alpha$ 234:
Giannini et al. 1999). 
ISO-LWS observations towards a sample of 11 Herbig Ae/Be stars shows that 
the OH lines are only detected towards the stars associated with 
with the largest mean densities (n$>$ 10$^5$ cm$^{-3}$) as derived from the 
OI(63 $\mic$)/CII(157$\mic$) ratio. The non-detection of OH in this nebula
is consistent with the idea that the bulk of the molecular gas is 
formed by the low density component traced by the 
OI(63$\mic$)/CII(157.7$\mic$) ratio.

A similar situation is found for CH and CH$_2$. Large CH
column densities have been detected in absorption. 
The data by Federman et al. (1997) shows that the CH column density 
in front of HD 200775 is 3.2$\pm$0.2 10$^{13}$ cm$^{-2}$. Even
if we assume excitation temperatures as low as T$_{ex}$ = 20 K, 
we should have detected
intensities of the order of 10$^{-9}$ erg s$^{-1}$ cm$^{-2}$ for the 180.93 and
149.09 $\mic$ lines. The non-detection of these lines can only be explained
invoking the excitation conditions. The beam filling factor of the
dense gas (n$>$ 10$^5$ cm$^{-3}$) is very low. 
Another possibility is that the CH abundance decreases for high
densities. This would also explain the lack of CH emission in other
sources. In fact, far to our knowledge,
emission of CH has not been detected towards any source
(just a tentative detection towards NGC 7027). 

\section{Summary: A model for the nebula}

In this paper we present the ISO-SWS and LWS observations towards 
a strip crossing the star and the most intense PDRs in the
reflection nebula NGC 7023. These observations altogether with
previous maps in molecular lines and the [HI] 21cm line have
allowed to figure out a model of the nebula.

We can consider the nebula as composed by three different regions. 
The first one is formed by the star, the small HII region around 
it and the low density gas filling the cavity. 
Federman et al. (1997) based on studies of absorption lines derived
that the density of the gas in front of the star is of a few 
10$^{2}$ cm$^{-3}$. A slightly larger value,
n = 10$^{3}$ cm$^{-3}$, was found by Gerin et al. (1998) based
on observations of the CO rotational lines towards the star. 
The Br$\alpha$ and [SiII] 34.8 $\mu$m lines peak 
in this region but the [HI] 21cm and H$_2$ rotational lines have a 
minimum in emission. The Br$\alpha$ emission is well explained as
arising in the stellar ionized wind modeled by Nisini et al. (1995).
However, the contribution of the
stellar wind to the intensity of the [SiII] 34.8 $\mu$m line is
expected to be negligible. The [SiII] 34.8 $\mu$m line is better
explained as arising in the diffuse gas filling the cavity in which
at least 20\% -- 30\% of the silicon is in gas phase.

The second region is formed by the walls
of the cavity in which the star is immersed. Clumpy PDRs have been
formed in these walls with high density filaments 
(n $\sim$ 10$^{5}$ -- 10$^{6}$ cm$^{-3}$) immersed in a more
diffuse medium( n $\sim$ 10$^{4}$ cm$^{-3}$). These high density
filaments are located almost symmetrically $\sim$ 50$''$ NW and 
$\sim$ 70$''$ SW (NW PDR and SW PDR) from 
the star and constitute the peaks of   
the [CII] 157.7 $\mu$m, [OI] 63.2 and 145.6 $\mu$m, [HI] 21cm 
and  H$_2$ rotational 
lines emission. 
The NW and SW PDRs have very similar excitation conditions 
although the SW PDR is a factor of 2 weaker than the NW PDR. This difference
in intensity is very likely due to a different beam filling factor.
In both, the NW and SW PDRs, the intensities of the H$_2$ rotational lines
can only be fitted assuming an ortho-to-para-H$_2$ ratio lower than 3.
Since the rotation temperatures of the H$_2$ lines ranges from 400 to 700 K,
this implies the existence of a non-equilibrium OTP ratio in the region.
The comparison between the vibrational and pure rotational H$_2$ lines
suggests that the OTP ratio increases across the PDR
from values as low as $\sim$ 1.5 for gas with temperatures 
$\sim$ 400 -- 700 K to
values close to 3 in the gas located at a visual extinction 
A$_v$ $<$ 0.7 mag (T$_k$ $\gsim$ 700 K).
This non-equilibrium OTP ratio has been interpreted 
in terms of an advancing photodissociation front.
The intensity of the [SiII] 34.8 $\mu$m line observed towards the NW PDR
is consistent with the predictions of PDR models with a standard silicon
depletion of $\delta$ Si = -1.3, i.e., only 5\% of the silicon is in gas
phase in these PDRs.  

The third region is the molecular gas. 
We have not detected the OH, CH and CH$_2$ molecular lines towards
this source. This is consistent with the weakness of these lines in other
sources and reveals that the beam filling factor of the 
dense gas (n $\geq$ 10$^5$ cm$^{-3}$) 
in the LWS aperture is low. The CO J=17$\rightarrow$16 line has
tentatively been detected towards the star.   

\acknowledgements{We would like to thank the referee F. Bertoldi
for his helpful comments and suggestions. We are also grateful to
J. Black for fruitful discussions on the chemistry of this region. 
This work has been partially supported by the Spanish
DGES under grant number PB96-0104 and the Spanish PNIE under
grant number ESP97-1490-E.
N.J.R-F  acknowledges Consejer{\i}a de Educaci\'on y Cultura de la
Comunidad de Madrid for a pre-doctoral fellowship.
}

\begin{figure*}[p]
\vspace{20cm}
\includegraphics{9257.f1}
\caption{Scheme of the observed region. The HI total column density
map (contours) is superposed on the integrated intensity map of the
$^{13}$CO J=1$\rightarrow$0 map (grey scale) (Fuente
et al. 1998). The filaments observed in H$_2$ fluorescent emission 
are also shown (Lemaire et al. 1996). The observed positions are indicated 
by filled squares and the star.}
\end{figure*}

\begin{figure*}[p]
\vspace{22cm}
\includegraphics{9257.f2}
\caption{ Spectra of the H$_2$ S(0),S(1),S(2),S(3),S(4)
and S(5) rotational lines , the SiII (34.8 \mic) line and the
Br$\alpha$ HI recombination line towards the five positions
observed across the PDR associated with NGC 7023 (see Fig. 1).
}
\end{figure*}

\begin{figure*}[o]
\vspace{9cm}
\includegraphics{9257.f3}
\caption{Normalized intensity of H$_2$ S(1), HI 21cm and SiII(34.8 $\mic$) 
lines as a function of the
distance from the star. Note that both, the H$_2$ and HI emission peaks
towards the H$_2$ filaments, while the SiII(34.8 $\mic$) 
emission peaks towards the star.    
}

\end{figure*}

\begin{figure*}[p]
\vspace{15cm}
\includegraphics{9257.f4}
\caption{ LWS spectra towards the NW PDR and the star position after
subtracting the continuum emission.
}
\end{figure*}

\begin{figure*}[p]
\vspace{20cm}
\includegraphics{9257.f5}
\caption{Rotational diagram of the pure H$_2$ rotational lines observed
towards the SW PDR. The points corresponding to para-levels are joined
by a dashed line, and the points corresponding to ortho-levels with a
thick line. Note that there is a shift between the ortho- and para-H$_2$
lines that is larger that the observational errors.
}
\end{figure*}

\begin{figure*}
\vspace{20cm}
\includegraphics{9257.f6}
\caption{Comparison of the observational data with the results of PDR
models with G$_0$ = 10$^4$, n = 10$^6$ cm$^{-3}$, and OTP ratios of
3 (open triangles), 1.5 (open circles) and 1 (open squares). The data are shown with
black squares and errorbars. The 
beam filling factor has been taken 0.5 for all the models. The model with
OTP = 1.5 fits the observational data.   
}
\end{figure*}

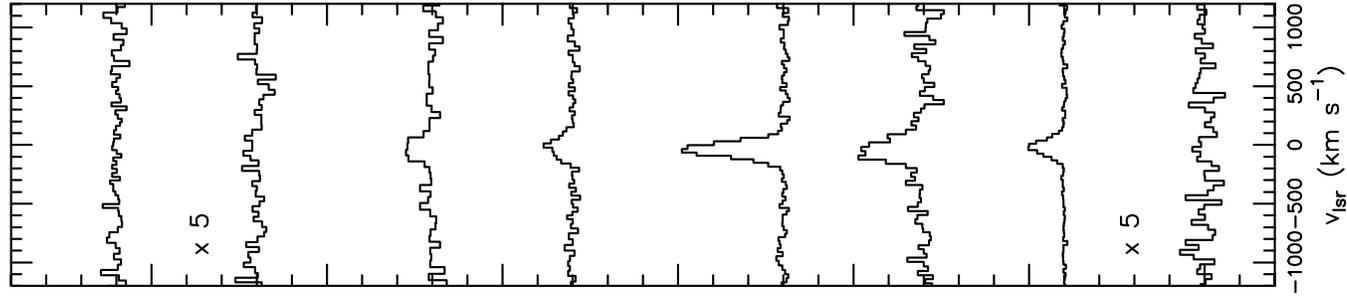
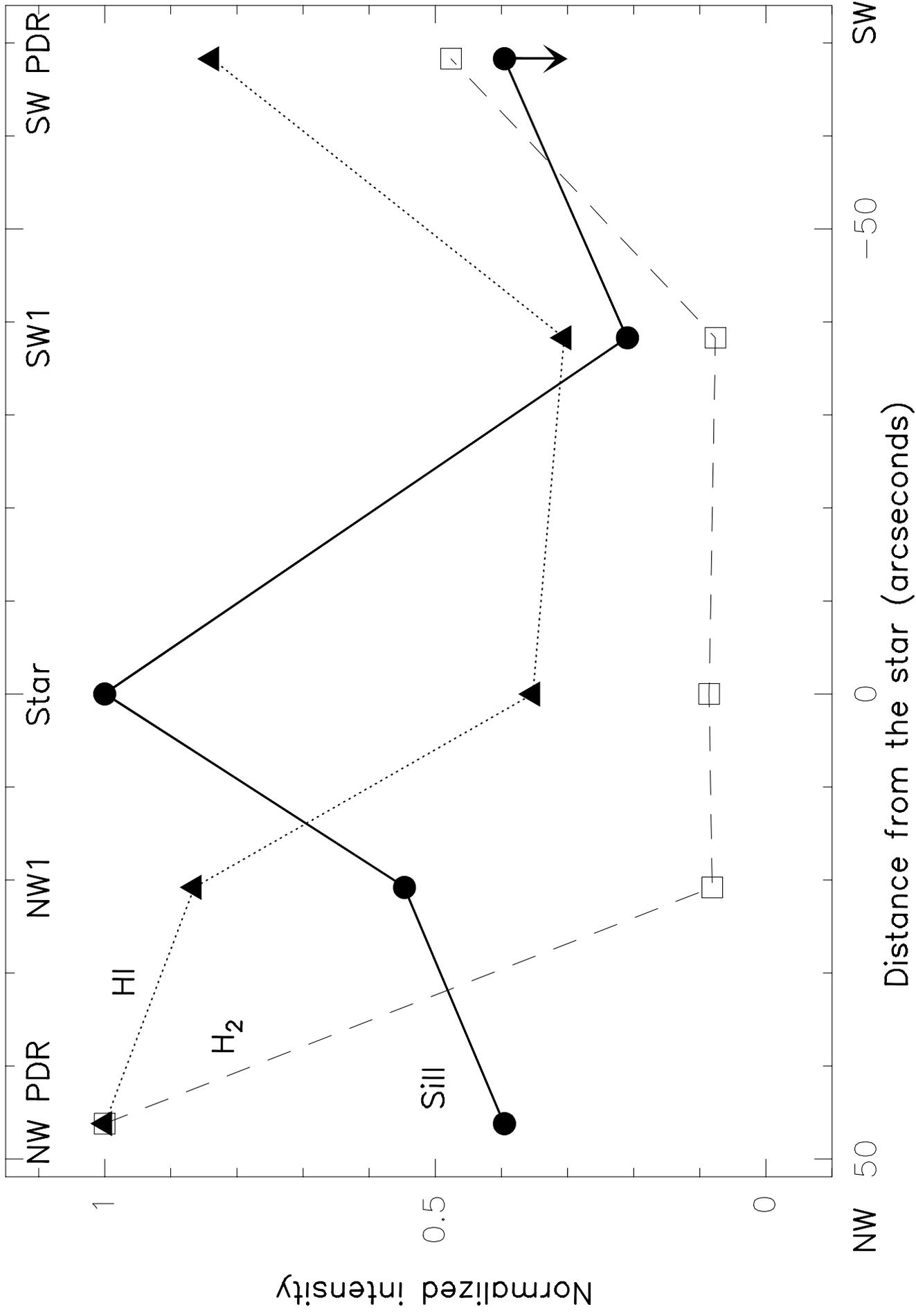
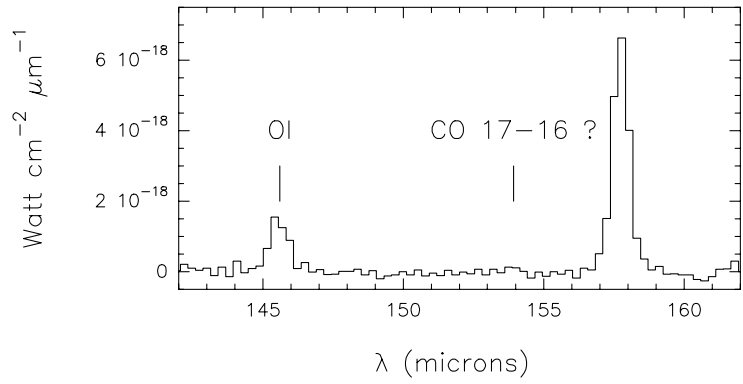
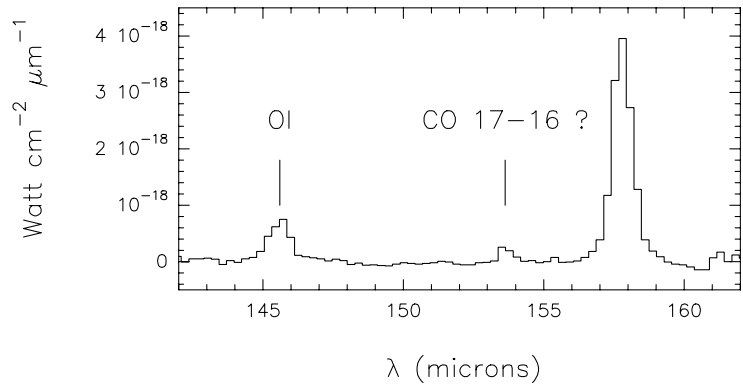
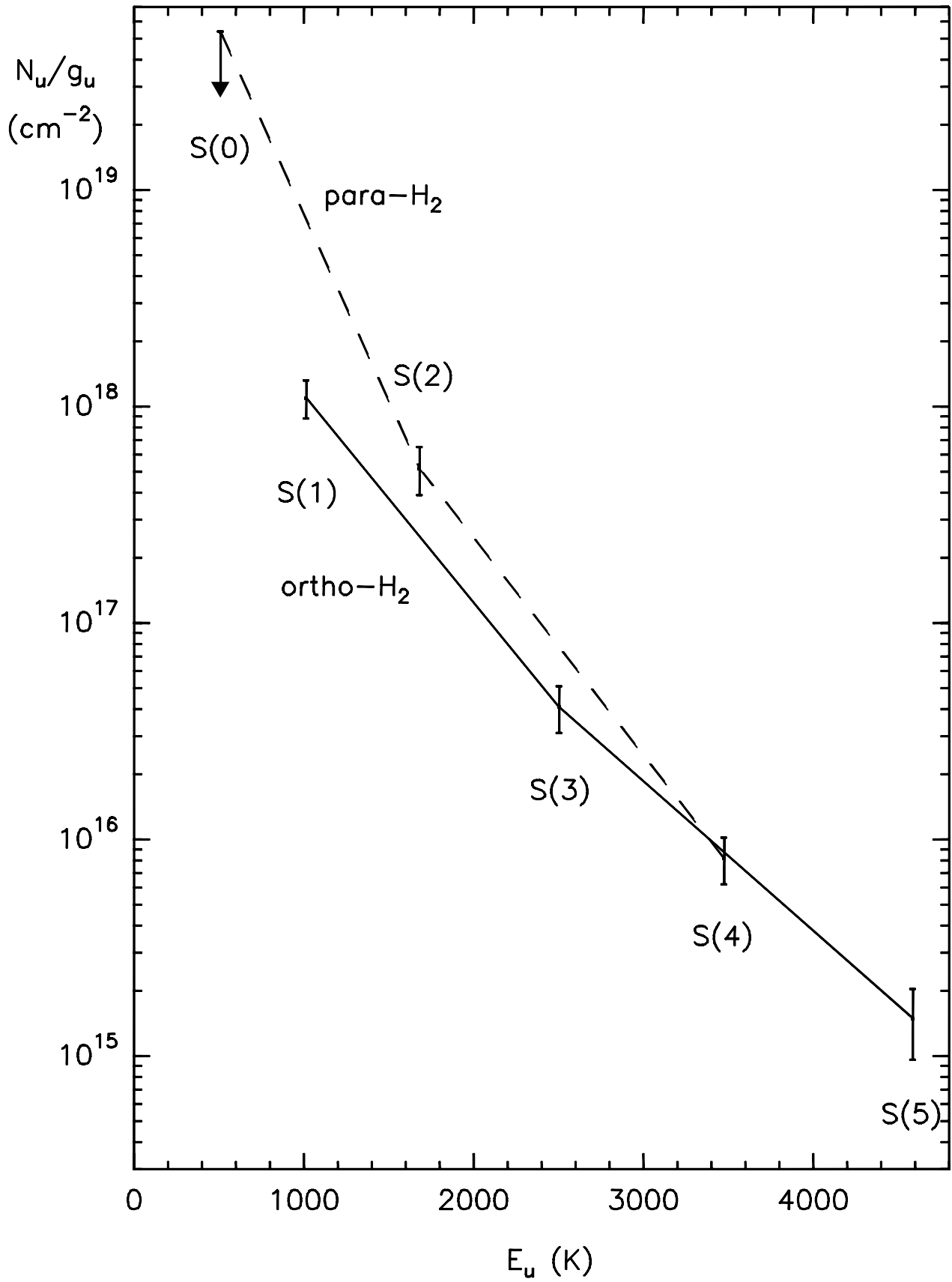
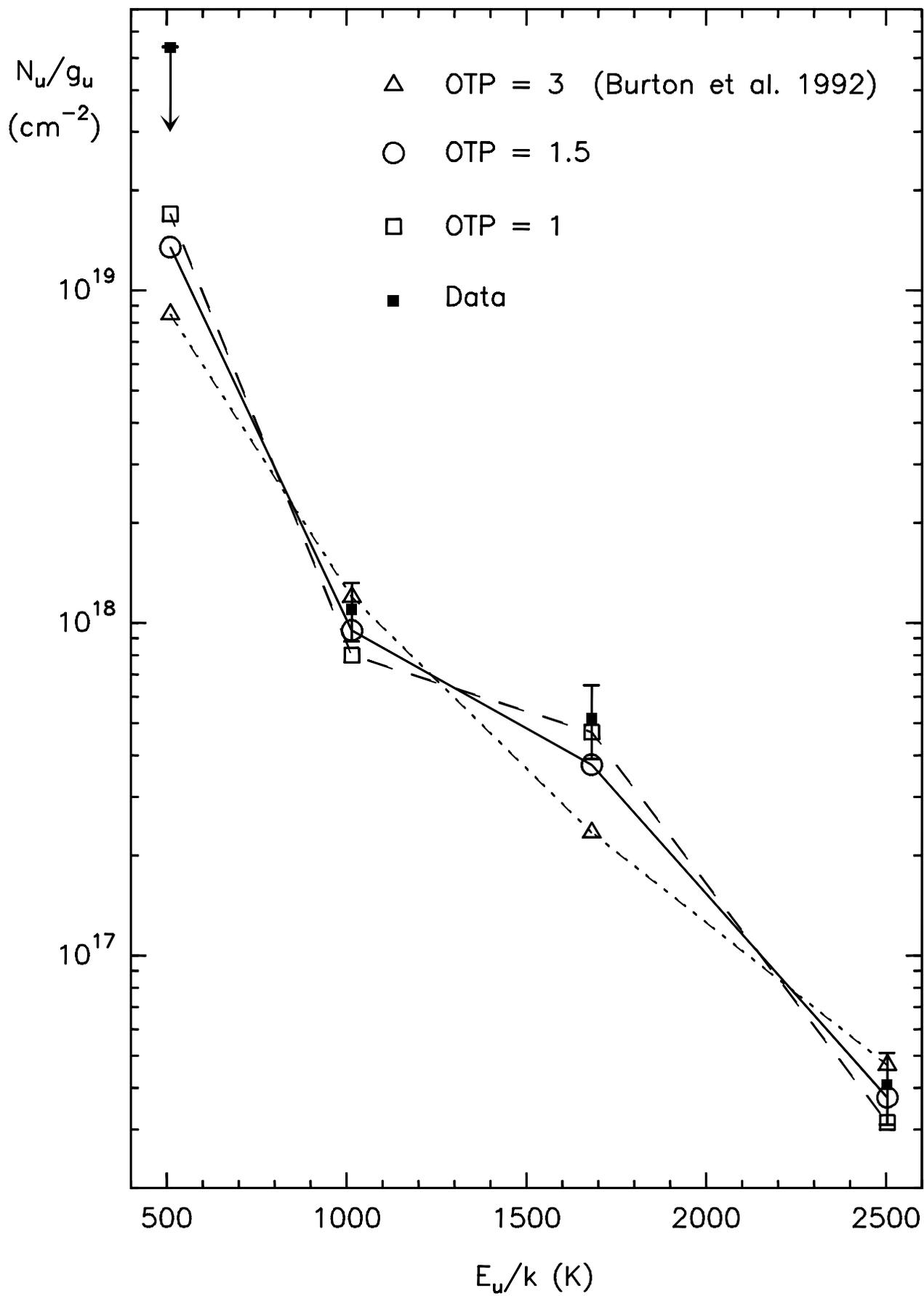
\end{document}